# Three Dimensionial Surface Modelling: A Novel Analysis Technique for Non-Destructive X-Ray Diffraction Imaging of Semiconductor Die Warpage & Strain in Fully Encapsulated Integrated Circuits.

J. Stopford [(1)(5)$], A. Henry[(2)], D. Allen[(1)], N. Bennett[(1)], D. Manessis[(3)], L. Boettcher [(3)], J. Wittge[(4)], A.N. Danilewsky [(4)] and P.J. McNally [(1)%]

(1) School of Electronic Engineering, Dublin City University, Dublin 9, Ireland.
(2) Department of Manufacturing Engineering, School of Manufacturing and Design Engineering, Dublin Institute of Technology, Ireland.
(3) Fraunhofer IZM Berlin, Microperipheric Research Center, Berlin Center of Advanced Packaging, Gustav-Meyer-Allee 25, Berlin, Germany.
(4) Albert-Ludwigs Universität Freiburg, Kristallographie, Institut für Geowissenschaften, Freiburg, Germany.
(5) Intel Ireland, Leixlip, Co. Kildare, Ireland.
$ *Jennifer.Stopford@Intel.com*
% *patrick.mcnally@dcu.ie*

**Abstract**

Future complementary metal oxide semiconductor (CMOS) scaling for advanced integrated circuit (IC) technologies may well depend on "More than Moore" (MtM) approaches using heterogeneous integration of semiconductor-based devices. Examples include the stacking of multiple silicon, or equivalent, die for radio frequency, analogue, digital, power and mixed-signal functions, etc. In order to realise this, advanced packaging technologies including System in Package (SiP), System on Chip (SoC) and 3D Integrated Circuits (3D ICs) are key enabling technologies [1]. However, these advanced packages are plagued by reliability problems and to date there is no proven or accepted non-destructive metrology which can simultaneously probe materials properties such as strain, warpage, dislocation generation, etc. in these systems from bare silicon die through to a fully encapsulated packaged system [1-6]. This problem still remains a major roadmap challenge [1]. We report herein on the development of a novel, x-ray diffraction amd analysis technique, which can address this major metrology gap, and we demonstrate the non-destructive production of X-Y spatial maps of deformations and strain fields in Si die inside fully encapsulated integrated circuit packages. The technique, which we call 3-dimensional surface modelling (3DSM), is used to obtain high resolution (~3 μm) strain/warpage maps, and quantitative information on the nature and extent of warpage in a demonstration quad no flat lead (QFN) advanced package running from early stage silicon die bonding through to the end of the manufacturing process, i.e. a fully encapsulated and production ready chip. 3DSM mapping produces heretofore unobservable imaging of the locations of major warpage features, encapsulation voiding, including their specific locations across the packaged chip and quantitative data on induced local wafer warpage all of which, crucially, have been acquired non-destructively and in *situ*. No pre-treatment or preparation of the samples are required These 3DSM maps can be acquired in a few minutes at an x-ray synchrotron source and provide for an accurate determination of the impact of individual process steps on wafer die warpage during the complex materials processing required for the production of next generation integrated circuit technology.

## 1. Introduction

The production of advanced MtM IC packages is hugely challenging as it represents the convergence of many materials challenges on to a very complex manufacturing process. Chief among these problems are the need to thin the semiconductor wafer (typically silicon) down to thicknesses of the order of 50 μm, the need to align accurately multiple thin die as they are stacked upon each other, the use of bond adhesives and glues, metallic bonding comprising typically of copper and tin alloys and the use of Through Silicon Vias (TSVs) in order to provide for direct communication between stacked silicon die [1-9]. This challenging convergence of nano- and macro-scale physics, materials science and engineering imposes many requirements for materials metrology including the requirement that, ideally, all metrology should be non-invasive, non-destructive and *in situ* to allow the real-time dynamical evaluation of package strains, warpage, defects, etc. which can then be improved through appropriate engineering intervention.

To date test metrologies for embedded chips have typically consisted of finite element modelling (FEM), micro-Raman spectroscopy, optical inspection, electrical testing and destructive testing such as shear testing [3-7]. Significant causes of strain, and hence failure of embedded chips are known to come from 3 major sources: materials induced stresses caused by differences in the coefficients of thermal expansion (CTE), wafer/chip warpage, and stress from the embedding/lamination processes [3-8]. Materials induced failures are predominantly located at or near through silicon vias (TSV), where differences in the CTE between the Si substrate and adjacent vias can lead to the formation of stresses in the surrounding substrate[5-6,9]. Wafer/die warpage increases greatly at thicknesses less than 100 μm, and bonding to the Cu substrate and subsequent lamination has been found to decrease warpage and relieve stresses. During lamination pressure and heat should be sufficiently high so as to provide sufficient adhesion and prevent voids occurring; however excess pressure and heat can lead to the formation of voids and delamination [2,7]. With so many diverse challenges to the realisation of high volume next-generation package integration, advanced non-destructive testing methodologies are fundamental to the development of embedded chip design and manufacture.

The work described in this letter is underpinned by a new approach to x-ray diffraction imaging (XRDI), also known as x-ray topography. XRDI in general is well understood and modelled by x-ray kinematical and dynamical diffraction theories [10,11]. In the laboratory it has been developed by the traditional Lang x-ray topographic method [12] while in 1974, Tuomi *et al.* pioneered the use of synchrotrons to produce high resolution synchrotron x-ray topographs with rapid throughput [13]. Due to the historically thick metallization layers and outer packaging, XRDI has rarely been used to analyse packaged or processed integrated circuits. The earliest attempts were on delidded packages [14], but notwithstanding the obvious fact that the package had to be delidded, the technique showed some early promise for the analysis of wafer warpage and non-destructive imaging of damage beneath wire bonds. More success has been achieved in the examination of the strain fields due to under bump metallisation (UBM), ball grid array solder bumps and on the impact of UBM on solder bump reliability [15-19]. However in all the aforementioned cases the packages, devices and/or bonds under test could not be examined in their actual processed state and the samples had to be prepared for scanning electron microscopy, thinned down for XRDI through the wafer back sides, and techniques such as micro-Raman spectroscopy could not "see" below metallisation in any case. The impending manufacture of relatively thinner packages required for MtM applications provided the opportunity to develop the 3DSM technique, whose efficacy is demonstrated herein.

## 2. Experimental

The packages examined in this study are QFN-A packages [2], measuring 160 μm thick and 10 mm × 10 mm in size. The package consists of an active die bonded Si chip, 5 mm × 5 mm in size and 50 μm thick, with a peripheral bond pad pitch of 100 μm, embedded face up in a substrate. The chip is covered from the top side with a resin-coated-copper (RCC) dielectric layer 90–100 μm thick. By way of proving the concept, the x-ray 3DSM analysis was performed at Stages (1)-(7) of the advanced package fabrication, and this letter will focus on the crucial exemplary Stages (1), (4) & (5) (See Appendix A).

XRDI was performed at the ANKA Synchrotron, Karlsruhe, Germany, using the TopoTomo beamline (see Appendix B). Section Transmission (ST) x-ray topographs were typically taken at 0.5 mm steps from the pads at the top of the chip stepping sequentially until the bottom of the chip, is reached. The original ST topograph images are scaled and cropped so that the topograph fills the complete frame. A set of 9 sketch planes parallel to a reference plane (right plane) are created in a well-known CAD package, SolidWorks®. Each sketch plane is positioned so that the distance between adjacent sketch planes is proportional to the distance between two corresponding ST topographs. Horizontal ST topographs are imported into SolidWorks® and positioned on the corresponding sketch planes so as to be centred with respect to the perpendicular front and side planes (Figure 1(a)).

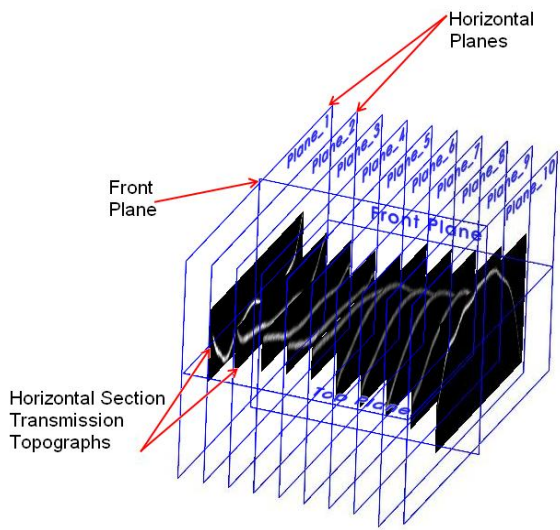 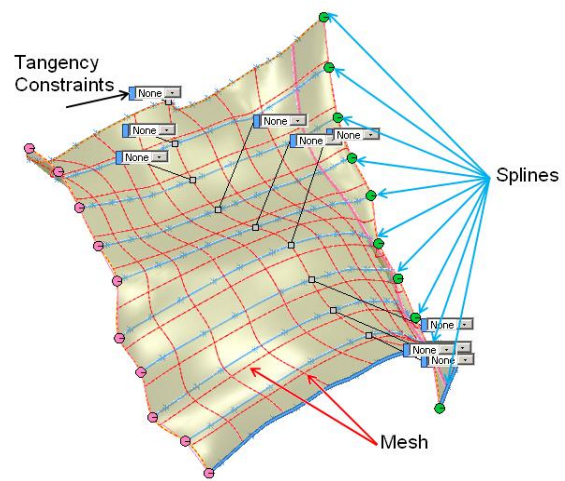

**Figure 1(a):** Construction of 3DSM showing 10 horizontal sketch planes with ST topographs intersected by front and top planes.

**Figure 1(b):** Boundary surface formation showing splines and mesh detail.

The next step in the modelling process is to create splines. SolidWorks® uses B-spline curves to define high resolution curves. B-spline curves are extended versions of Bezier curves, with each B-spline consisting of a number of Bezier curve segments. B-spline curves are made up of an arbitrary number of control points which provide a basis for the contour to approximate. In order to model the shape of the imported ST topographs the contours of the topographs are followed by positioning points along the exterior contours, and using the point option of the spline mode in SolidWorks®. B-spline curves are filled to the points using the curve fitting technique, which utilises the least square fitting method. The B-spline curve is then adjusted further so as to fit the contours of the topograph more closely. This is carried out by adjusting the handles which control the tangent vectors at the spline points. B-spline curves are created for each of the typically 10 ST topographs acquired for these proof-of-concept tests. The imported topographs are then hidden, enabling a set of splines to be displayed. The 3D surface model is created using the boundary surface feature in SolidWorks®. The boundary surface feature creates a solid surface model by connecting the splines created in the previous step. The splines are selected in sequence, from the P1 horizontal spline to the P10 horizontal spline, ensuring that the same ends of the splines are selected to ensure surface continuity. Figure 1(b) shows the boundary surface mesh across 10 horizontal splines; no tangency constraint, i.e. zero curvature is applied in order to give the closest possible fit of the surface to the splines. The surface can also be created by lofting between splines. However boundary surfaces are generally considered of higher quality as the surface created is continuous in all directions.

### 3. Analysis of QFN Packages

*Stage 1: Chip Attach*
The first stage in the embedding process is chip attach. Chips are initially prepared by electrolytic deposition of 6-8 μm of Cu to the bond pads. Using a high precision (± 10 μm) die attach machine, double layered die attach film (DAF), an adhesive layer 20 μm thick is used to bond the silicon chip face-up to the Cu substrate (Figure 2(a)).

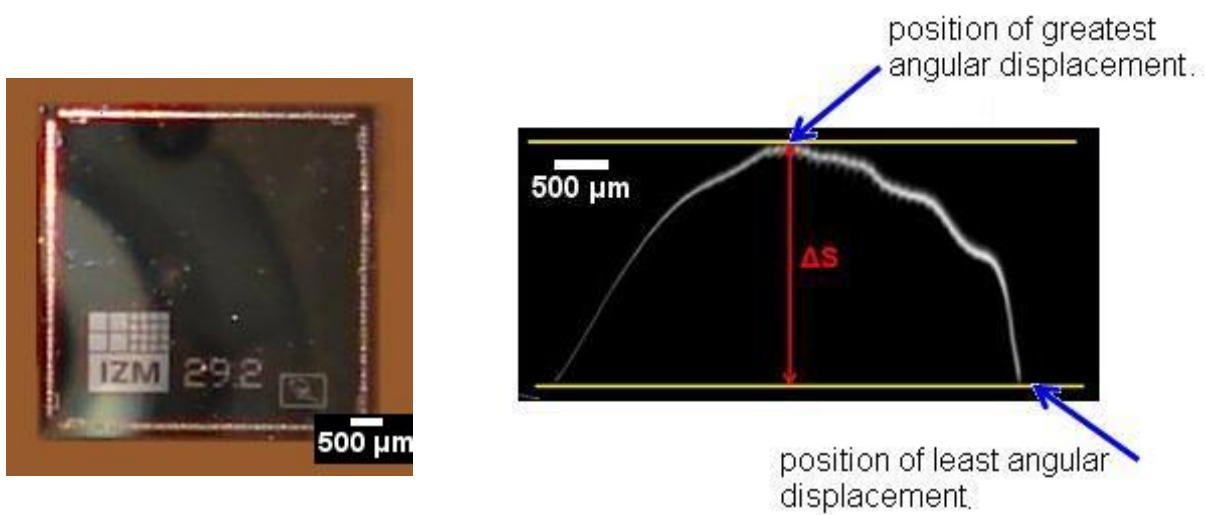

**Figure 2(a):** Unpackaged, but back   **Figure 2(b):** 2 2 0 ST topograph of Stage 1 sample, This side bonded, chip from QFN-A package at Stage 1 of the embedding process.

topograph is number 1 in series of 10 ST topographs which are processed to form images of Figs. 2(c)-(d). Yellow lines show maximum angular deviation across the ST topograph, which is equal to ΔS.

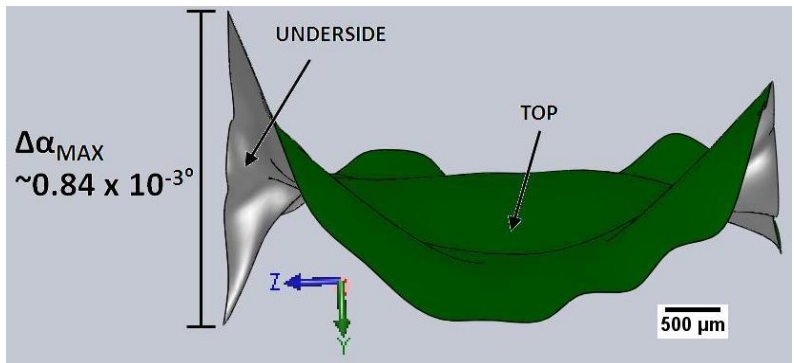

**Figure 2(c):** 3DSM map showing Δα, the relative misorientation of (2 2 0) Si planes of Stage 1 sample viewed end on, with ST position corresponding to the upper edge of the chip facing the page.

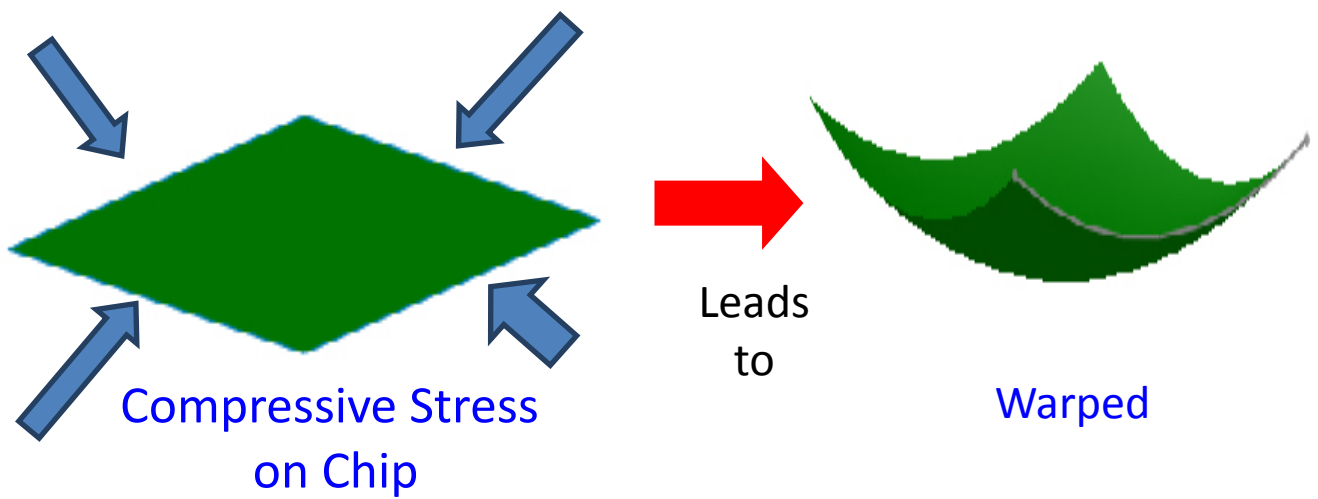

**Figure 2(d):** The shape of the induced wafer warpage derived from micro-Raman spectroscopy measurements. The overall wafer curvature is similar to that of the 3DSM data of Figure 2(c) although the 3DSM data shows much greater local resolution.

The 3DSM maps are created from a series of up to 10 ST topographs as described previously. In unstrained Si these ST topographs would consist of a series of straight "slice images" through the Si running from the top surface through to the back surface. However, the severe distortion in the individual ST topographs (the straight line is now very "wiggly") is formed by the orientational contrast mechanism, whereby diffracted x-rays from highly strained and hence misorientated, regions in the Si lattice possess an altered Bragg angle. Essentially each topograph is a spatial record of the variation of the Bragg angle of the Silicon across the line of the section topograph, in this case across the entire 5mm width of the chip. For the orientational contrast to be recorded, a good approximation is that the lattice misorientation must exceed the full width at half maximum of the x-ray beam divergence, which is a maximum of 2 mrad in the case of the ANKA Synchrotron [21]. The magnitude of the recorded shift due to this lattice misorientation, or warpage is approximated by:

$$\Delta S / 2L \cong \Delta \alpha \qquad (1)$$

where ΔS is the measured angular misorientation across the measured length of the sample, as shown in Figure 2(b), L is the sample the sample to film distance, and Δα is the maximum apparent shift in the Bragg angle due to the strain-induced tilt of the diffracting planes [16, 18, 22]. $\Delta\alpha_{MAX}$, the maximum misorientation of the Si planes for the reflection imaged using XRDI, is therefore measured relative to the points of greatest and least angular displacement in the relevant ST topograph. Figure 2(c) shows the 3DSM map for the Stage 1 sample. The greatest lattice distortion is observed towards the top of the chip, as seen in Figure 2(c). $\Delta\alpha_{MAX}$, the maximum {220} Si plane lattice tilt for this sample is $0.84 \times 10^{-3}$ (± 5%) degrees.

As a means of testing the validity of the technique, micro-Raman spectroscopy data were acquired across these unencapsulated Stage 1 samples using a λ = 488 nm Ar$^+$ laser, which has a penetration depth of ~550 nm in Si, and thus the data are indicative of strain levels close to the surface of the chip. This confirmed strain levels of the order of hundreds of MPa, with the greatest levels of compressive strain apparent at the edges and corners of the chip (~360 MPa), which suggests a chip warpage shape as shown in Figure 2(d). This overall curvature compares very well with the 3DSM data, which exhibit a uniform, relatively undistorted (warped) region at the centre of the chip, and significant warpage of the {2 2 0} lattice planes along all 4 edges of the chip, albeit with much resolved detail. Although the magnitude of the induced lattice warpage, as seen in the 3DSM, and measured from the ST topographs is relatively small at this early processing stage, micro-Raman spectroscopy data confirm that close to the surface, the induced strain levels are high and have the potential to affect device functionality and/or reliability [2, 6, 23]. Large area x-ray topographs (not shown here) cannot image effectively the strain field induced warpage in the chip, demonstrating that detailed information on lattice warpage is not always apparent from conventional XRDI, and that 3DSM is a valuable tool in characterisation of warpage/strain in packaged chips.

*Stage 4: Via Electroplating*
The second stage of the packaging process involves applying a 90–100 μm thick resin coated copper (RCC) layer to the chip from the top side using a standard printed circuit board (PCB) multilayer vacuum lamination process, curing the epoxy followed by drilling of micro-vias through to the chip pads at the edge of the Si die. After Stage 4 processing, the die is completely enclosed in Cu and no part of the die is visible (Figure 3(a)). Micro-Raman spectroscopy is not feasible beyond the first stage of the packaging process due to the presence of this Cu metallization. Likewise, none of the currently available IC characterisation metrologies can non-destructively measure or image stress/strain, warpage or defects beneath the package lid [24].

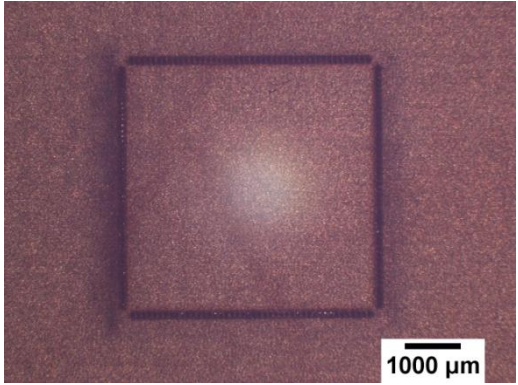

**Figure 3(a):** Optical micrograph of die after Stage 4, drilling of micro-vias to the Cu pads.

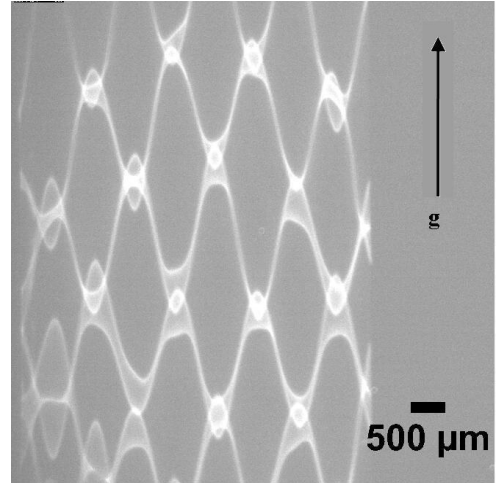

**Figure 3(b):** 2 2 0 LAT topograph taken at central section of stage 4 sample.

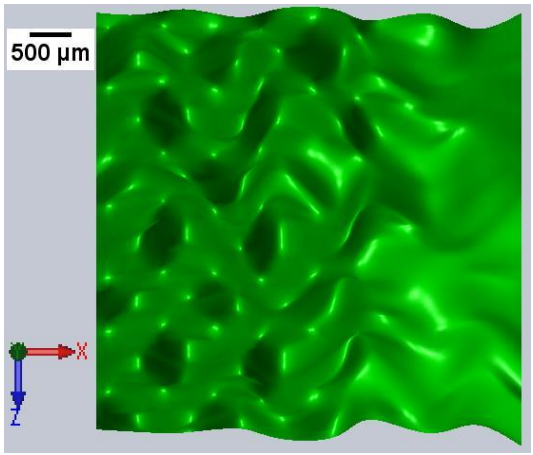

**Figure 3(c):** 3DSM plot of misorientation of (2 2 0) Si planes inside stage 4 sample.

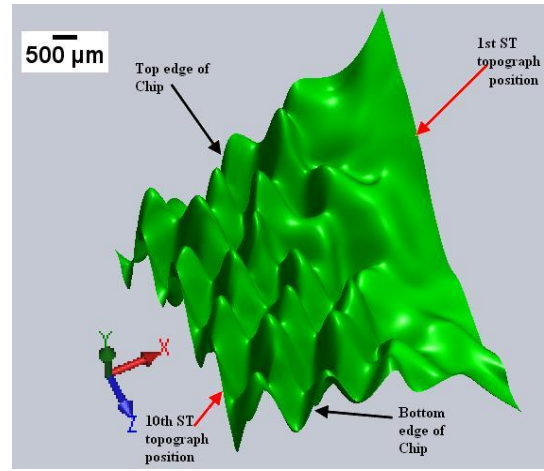

**Figure 3(d):** 3DSM plot of misorientation of (2 2 0) Si planes inside stage 4 sample.

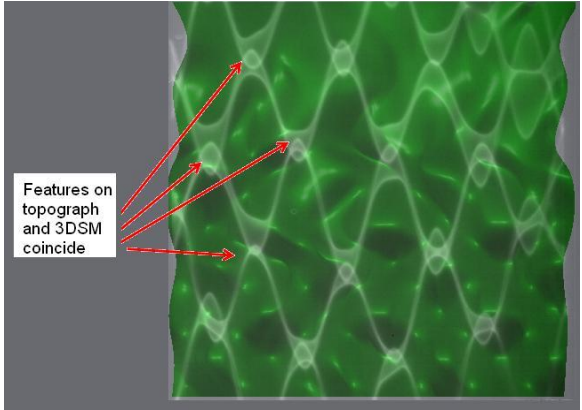

**Figure 3(e):** 3DSM of misorientation of (2 2 0) Si planes inside Stage 4 sample with LAT overlaid. Scale is same at that for Figure 3(b).

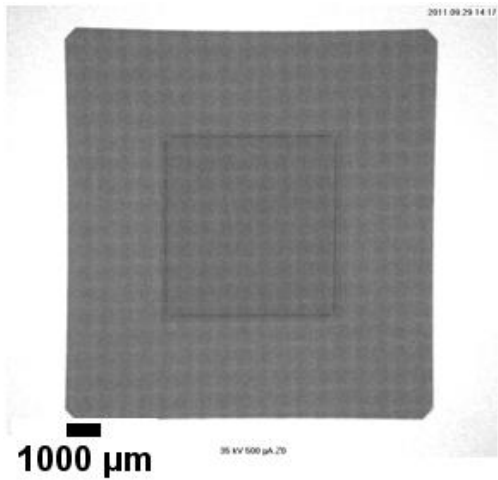

**Figure 3(f):** X-ray radiograph image of Stage 4 package showing embedded chip. Image is captured on a *Phoenix PCB Analyser.*

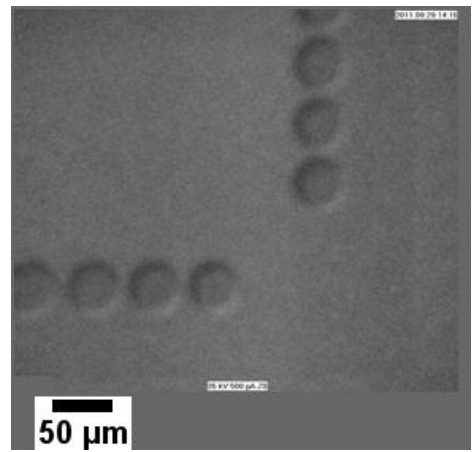

**Figure 3(g):** X-ray radiograph of Stage 4 package showing TSVs. Image is captured on a *Phoenix PCB Analyser.*

Large area transmission (LAT) topographs of the Stage 4 samples display a distinctive wave pattern, with a peak to peak pitch of 1050 – 1150 µm. The wave pattern is most prominent in the central region of the chip, with peaks smoothing out and becoming less pronounced towards the edges of the chip. Initial examination of the LAT topographs in Figure 2(b) gives little information on the nature of the lattice distortion which would give rise to such a pattern. However the 3DSM plots clearly show the pattern of strain induced die warpage inside the packages (Figures 3(c) & (d)). Figure 3(e) illustrates the alignment of the 3DSM data to the wave-like pattern on the LAT topograph, and further demonstrates the efficacy of the 3DSM process for encapsulated packaged chips. $\Delta\alpha_{MAX}$, the relative angular lattice misorientations across the die is ~0.65-0.9 × $10^{-3}$ (± 5%) degrees for this sample.

The development of the regular pattern of bumps/peaks at this stage in the process, and its absence at Stage 1 suggests the phenomenon is linked to either the chip embedding by vacuum lamination or the via drilling processes. The lamination process for QFN-A packages has undergone comprehensive development and optimisation over the course of several projects [25-27]. Heating rates and pressure are critical parameters with respect to lamination integrity. High heating rates at low pressures (5–10 bar) can lead to the occurrence of voids, and excessive epoxy fluid viscosity has the potential to cause insufficient epoxy coverage around the chip and deleteriously influence package flatness [2-3]. The distinctive pattern of lattice warpage seen after Stage 4 processing is therefore most likely linked to the vacuum lamination process, where the presence of voids or variations in epoxy coverage may lead to localised variations in strain in the crystal lattice, resulting in the formation of the bump-like structures of Figures 3(b)-(e). While this is the subject of ongoing investigations, the important thing to note is that the 3DSM methodology allows us to analyse the unique impact each critical process step, which has heretofore not been possible. Finally, Figures 3(f)-(g) show x-ray radiograph images of the Stage 4 chip captured on a *Phoenix PCB Analyser*, a popular tool in the semiconductor industry for PCB and packaged IC failure analysis. The fibreglass weave pattern of the PCB substrate, and the dimpled appearance of the TSV surrounding the embedded chip can be clearly identified. However, when we compare these to the 3DSM data, one can observe that they are incapable of resolving strain and/or warpage inside the chip packages.

*Stage 5: Post Production*
The final example outlined is a Stage 5 sample defined as a post production sample. Figure 4(a) shows a large area transmission (LAT) topograph of the sample, the region circled in red corresponding to a region of major lattice distortion.

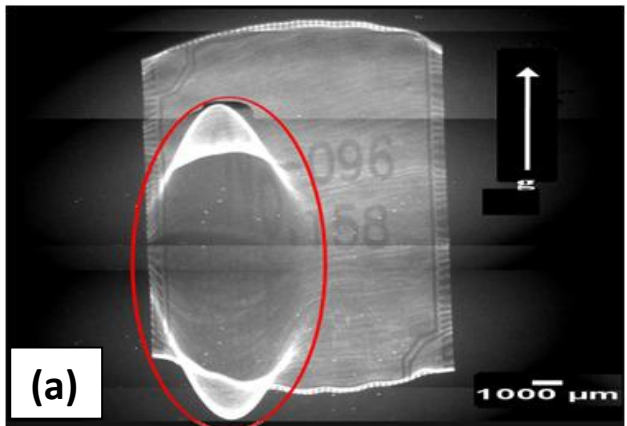

**Figure 4(a**): LAT topographs of stage 5 sample. 220 reflections were recorded using a CCD camera, and images are patched together to show entire chip. Region circled in red corresponds to region of high lattice distortion.

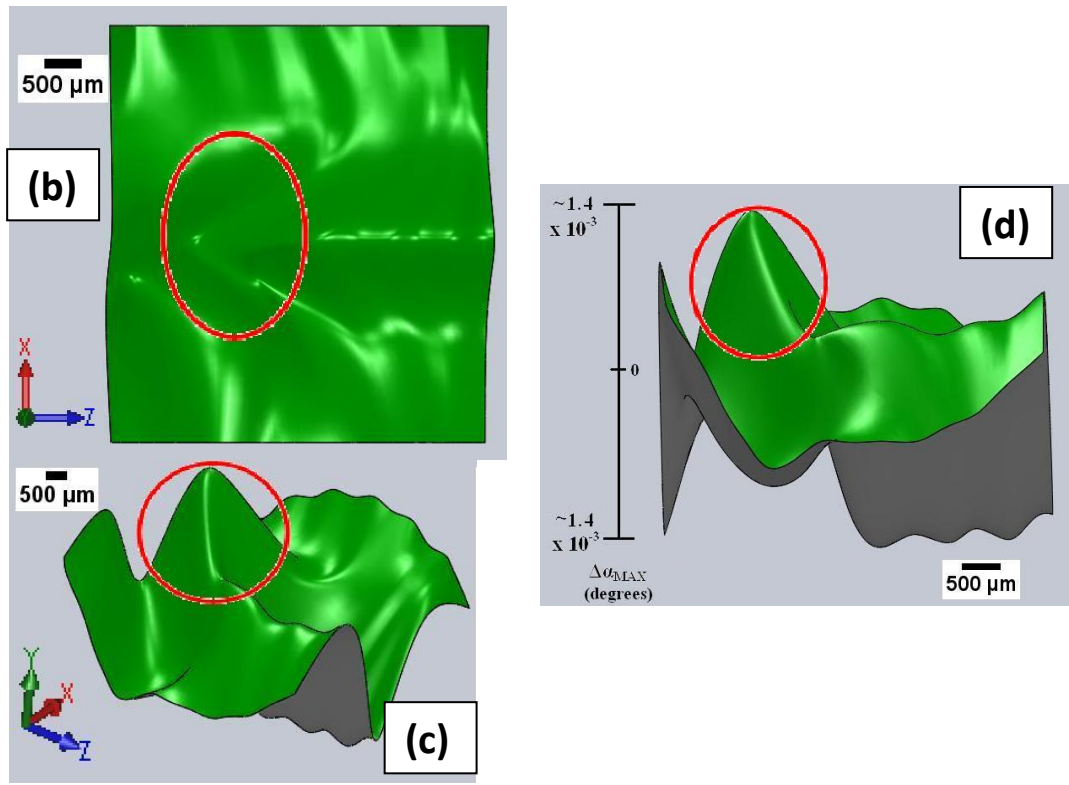

**Figures 4 (b), (c), (d):** 3DSM maps of misorientation of (2 2 0) Si planes inside a completely sealed stage 5 package. Red circles correspond to the elliptical region in Figure 4 (a).

While the conventional LAT topograph is capable of locating the position of this distorted region it is quite incapable of providing an analysis of the nature of this distortion. The 3DSM mapping illustrates in greater detail the nature of this distorted region. The magnitude of the measured angular misorientation ranges from ~$1.69 \times 10^{-3}$ (± 5%) degrees at the distorted region to ~$0.15 \times 10^{-3}$ (± 5%) degrees across the remainder of the chip.

Chip support and clamping in die bonding and chip placement processes are critical for die placement accuracy. When support is insufficient warpage can occur [27]. The consistently elliptical shape and the uniform nature (repeatable die warpage profiles in 3DSM as per Fig. 4(b)-(d)) were observed in numerous Stage 5 packages) suggest strongly that they are linked to the manufacturing process. They may be due to the supports or clamps in die bonding and chip placement processes, perhaps where a die or substrate is supported on two sides, wherein stress from processing can manifest itself in unsupported planes. The presence of vias can also lead to localised stresses in the package and TSVs can also induce stress by squeezing/stretching the adjacent material leading to material deformation, debonding and delamination [28]. The fact that 3DSM mapping can pinpoint non-destructively and *in situ* the precise location of the warpage/delamination problems on-chip will allow manufacturers to address these problems during production development or in-line with future developments of the technique.

### 4. Conclusion
In conclusion, there is no proven methodology for non-invasive and non-destructive stress and strain metrology in Systems on Chip, System in Package, 3D ICs and similar advanced packages as recognised in the International Technology Roadmap for Semiconductors. 3DSM mapping provides a unique ability to non-destructively assess strain/warpage inside sealed packages post production and after reliability testing, and this technique and its variants are applicable to many semiconductor, photonic and electronic package metrology problems for which there is currently no available solution. It is also conceivable that variants of this technique could be modified for in-laboratory and real-time applications. The authors are indeed investigating this possibility and this will be presented in a later paper.

**Acknowledgments**
P.M.N., J.S. and A.C. acknowledge the support of Science Foundation Ireland's Strategic Research Cluster Programme ("Precision" 08/SRC/I1411). N.B. and C.S.W. thank the Irish Higher Education Authority INSPIRE programme under the National Development Plan 2007-2013. J.S., P.M.N. and D.A. acknowledge the support of the project entitled "Smart Power Management in Home and Health" (SmartPM), EU FP7 ENIAC-2008-1 Joint Undertaking, Project # 120008 (Enterprise Ireland component # IR-2008-0010). This work was also supported by the European Community—Research Infrastructure Action under the FP7 ''European light sources activities - synchrotrons and free electron lasers (ELISA)'' programme.

**Author contributions**

# Appendix A – Alignment of Samples for XRDI

In order to apply the 3DSM technique the x-ray diffraction images (topographs) were recorded in section transmission (ST) geometry (Figure A1(a)). The beam was collimated into a narrow ribbon by a slit only 15 µm in height. Radiation passes through the back side of the sample, and reflections are recorded on high resolution Geola VRP-M film or a CCD. The narrow slit size enables sample cross-section images to be obtained, providing the Bragg angle is not too small. Digital topographs are recorded using a PCO.4000 high resolution 14 bit cooled CCD camera, with magnification optics, which enables a pixel size of 2.5 µm to be obtained [1]. The sample to camera distance, L, is set to 80 mm. The CCD camera is positioned to record the 220 reflection from the Laue diffraction pattern of topographs.

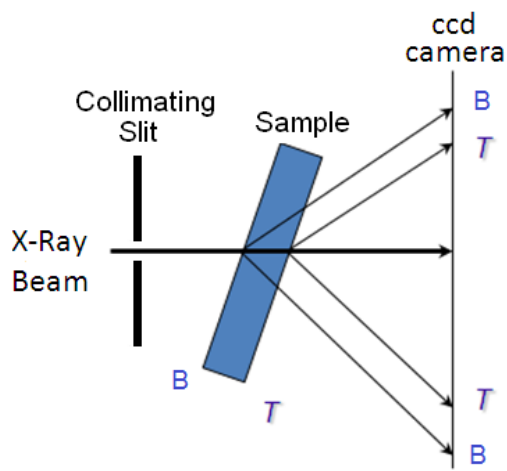
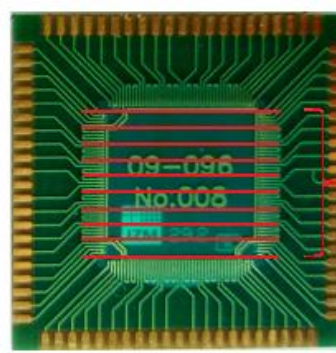

**Figure A1(a):** Section Transmission Topography. The sample is tilted to 12° to optimise the position of the 220 reflection. B and T represent the bottom and top of the sample respectively.

**Figure A1(b):** Optical micrograph of the QFN package showing the approximate positions where ST topographs where obtained.

The sample is mounted on a high precision X-Y stage which enables the user to step across the sample and thus obtain a series of section topographs across the packaged chip. ST topographs are taken at, typically, 0.5 mm steps from the pads at the top of the chip stepping sequentially until the bottom of the chip, is reached (Figure 1(b)).

# Appendix B - Details of QFN-A Package Fabrication

The packages examined in this study are QFN-A packages, measuring 160 μm thick and 10 mm × 10 mm in size. The package consists of an active die bonded Si chip, 5 mm × 5 mm in size and 50 μm thick, with a peripheral bond pad pitch of 100 μm, embedded face up in a substrate. The chip is covered from the top side with a resin-coated-copper (RCC) dielectric layer 90–100 μm thick. A schematic of the chip assembly, processing and testing steps is shown in Figure B1 . Prior to embedding, chips were prepared by electrolytic deposition of 6-8 μm of Cu to the bond pads. Using a high precision (± 10 μm) die attach machine, double layered die attach film (DAF), an adhesive layer 20 μm thick was used to bond the chip face-up to the Cu substrate (Stage (1)). A RCC layer 90 – 100 μm thick was applied to the chip from the top side using a standard printed circuit board (pcb) multilayer vacuum lamination process [2], and the epoxy was then cured at ~185 °C for 60 min (Stage (2)). A pulsed 355 nm UV laser was used to drill the microvias through to the chip pads (Stage (3)). The vias were then cleaned to remove excess epoxy resin and to improve Cu adhesion, and palladium was deposited on the epoxy surface prior to Cu metallization via electroplating (Stage (4)). Structured Cu conductor lines are required to connect the bond pads and capture pads on the chip. This was undertaken by means of laser direct imaging (LDI), which was used to expose a negative resist, which was subsequently acid etched to reveal the Cu line structure. The final step in the packaging process is Cu structuring on the bottom side of the package (Stage (5)). Packages were processed in large panel format, and separated by cutting or sawing [3] before undergoing pre-conditioning and qualification tests, Stages (6) & (7). These were in line with the JESD22-A113-E standard [4].

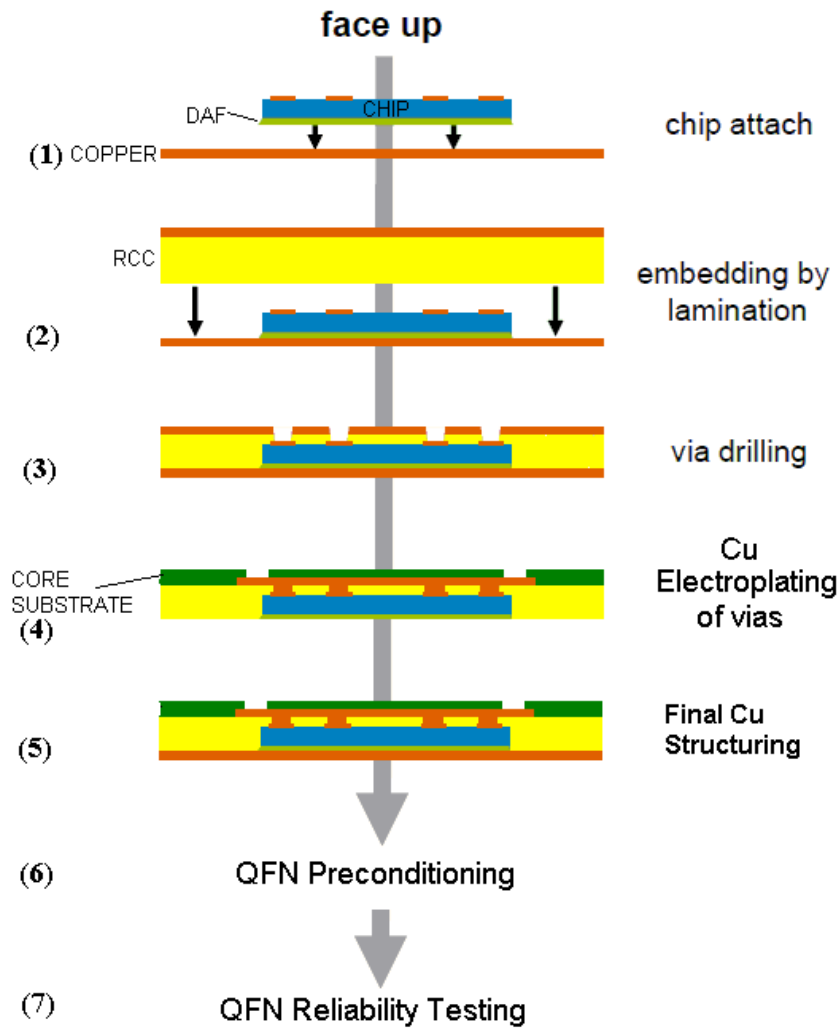

**Figure B1:** Schematic of major chip embedding stages for QFN-A package. The QFN-A chips also underwent preconditioning and reliability testing after assembly (Stages (6) & (7)).